\documentclass[11pt]{article}
\usepackage{graphicx}

\usepackage{graphicx}
\usepackage{dcolumn}
\usepackage{amsmath}
\usepackage{epsfig}
\usepackage{dcolumn}
\usepackage{natbib}
\usepackage{psfrag}

\newcommand{\BABARPubYear}    {08}

\newcommand{\BABARProcNumber} {066}
\newcommand{\SLACPubNumber} {13408}

\input babarsym

\long\def\inst#1{\par\nobreak\kern 4pt\nobreak
    {\it #1}\par\vskip 10pt plus 3pt minus 3pt}

\begin{document}
{\pagestyle{empty}

\begin{flushright}
SLAC-PUB-\SLACPubNumber \\
\babar-PROC-\BABARPubYear/\BABARProcNumber \\
September, 2008 \\
\end{flushright}

\par\vskip 4cm

\begin{center}
\Large \bf New Physics Searches at \lbabar
\end{center}
\bigskip

\begin{center}
\large 
F.~Renga\\
Universit\`{a} di Roma ``La Sapienza'' and INFN Roma \\
(for the BABAR Collaboration)
\end{center}
\bigskip \bigskip

\begin{center}
\large \bf Abstract
\end{center}
We will present the most recent results from the \babar\ Collaboration
concerning New Physics searches in rare $B$ and Lepton Flavour Violating
(LFV) decays, including $b \to s$ transitions, purely leptonic $B$ decays and
LFV $\tau$ decays.

\vfill
\begin{center}
Presented at the 16th International Conference On Supersymmetry And 
The Unification Of Fundamental Interactions (SUSY08)\\
06/16/2008---06/21/2008, Seoul, Korea
\end{center}

\vspace{1.0cm}
\begin{center}
{\em Stanford Linear Accelerator Center, Stanford University, 
Stanford, CA 94309} \\ \vspace{0.1cm}\hrule\vspace{0.1cm}
Work supported in part by Department of Energy contract DE-AC02-76SF00515.
\end{center}

\section{Introduction}

Rare $B$ and $\tau$ decays are a standard probe for New Physics (NP) searches
and a $B$-Factory provides a unique environment to look for these processes.
Many rare decays can be investigated and are sensitive to different NP
scenarios. 

At first, we can consider $B$ decays mediated by flavour-changing neutral
currents (FCNC). They are characterized by low rates in the Standard
Model (SM), due to the absence of FCNC at tree level. Hence, NP effects can be
of the same order of magnitude of the SM contributions. The high number
of $B \overline B$ pairs produced by a $B$-Factory often allows to approach
or reach the experimental sensitivity needed to observe these decays, and
strong constraints can be put on the NP contributions, by comparing the
experimental results with the SM expectations. 

We can also consider purely leptonic $B$ decays and LFV $\tau$ decays. The very
low SM rate of these decays often make them unaccessible with the present
experiments, unless NP effects enhance the rate up to the current experimental
sensitivity. For some of these decays, just the observation by itself would
provide an unambiguous evidence of NP.

In this work, we present the most recent NP searches in rare $B$ and
$\tau$ decays, based on the data collected by the \babar\ detector~\cite{NIM}
at PEP-II, an asymmetric $e^+e^-$ collider operating at a center of
mass energy of 10.58 $GeV$, corresponding to the mass of the $\Upsilon(4S)$
resonance.

\section{Electroweak Penguins}


The FCNC decays proceeding through electroweak penguin
diagrams show a good sensitivity to supersymmetric models and other NP
scenarios. 

Among them, the $B \to X_s \gamma$ process has been deeply
investigated both from the theoretical and experimental point of view in the
last years. A SM estimate of the inclusive branching ratio (BR) is available at
the NNLO~\cite{misiak}:
\begin{equation}
 \mathcal{B}(B \to X_s \gamma) = (3.15 \pm 0.23) \times 10^{-4}\, .
\end{equation}
The CP asymmetry:
\begin{equation}
 A_{CP} = \frac{\Gamma(\overline B \to X_s \gamma)-\Gamma(B \to X_{\overline s}
\gamma)}{\Gamma(\overline B \to X_s \gamma)+\Gamma(B \to X_{\overline s}
\gamma)} \, , 
\end{equation}
is also interesting, since it is expected to be at the level of 0.5\% in the SM
but can be strongly enhanced by NP effects~\cite{hurth}.

The \babar\ Collaboration produced several results for the inclusive BR,
adopting different techniques. The most stringent measurement, based on a sample
of about $88.9 \times 10^6$ $B \overline B$ pairs, has been done by means of a
semi-inclusive approach~\cite{babar_Xsgamma_semi} and gives:
\begin{equation}
 \mathcal{B}(B \to X_s \gamma) = (3.27 \pm 0.18 ^{+0.55}_{-0.41}) \times
10^{-4}\, ,
\end{equation}
where (here and in the following) the first error is statistical, the
second is systematic. In this analysis, 38 exclusive final states are
reconstructed (22 modes with 1 kaon and 1 to 4 pions, 10 modes with a kaon, an
$\eta$ and up to 2 pions, 6 modes with 1 kaon and up to 1 pion). The signal
yields are combined in order to extract the inclusive BR. The error is dominated
by the systematic uncertainty coming from the modeling of the parton
fragmentation and the contribution from unreconstructed modes. A similar
technique has been used in order to measure the CP asymmetry, obtaining:
\begin{equation}
 A_{CP} = -0.011 \pm 0.030 \pm 0.014 \, ,
\end{equation}
in a sample of about $383 \times 10^6$ $B \overline
B$ pairs~\cite{babar_Xsll_ACP}.

Recently, the Collaboration produced a new result for the BR, based on a recoil
technique~\cite{babar_Xsgamma_reco}. In this kind of analysis, one of the two
$B$ ($B_{tag}$) is reconstructed in a frequent decay mode, while the signal
signature is searched for in the rest of the event (the \emph{recoil}), composed
by all tracks and neutral particles not associated to the $B_{tag}$. This
technique provides a pure sample of $B \overline B$ events and a clean
environment to look for rare decays. 

The $B \to X_s \gamma$ BR is measured on the recoil of hadronic $B \to DX$
decays. More than 1000 exclusive $B_{tag}$ decay modes are
used. The kinematical consistency of the $B_{tag}$ is checked with the two
variables $m_{\mathrm{ES}} = \sqrt{E^{*2}_{\mathrm{beam}}-|\mathbf
p^{\,*}_{B}|^2}$ and $\Delta E = E^*_{B}-E^*_{\mathrm{beam}}$,
where $E^*_{\mathrm{beam}}$ is the beam energy and $E^{*}_{B}$ and
$\mathbf p^{\,*}_{B}$ are the energy and the momentum of the $B_{tag}$ in the
center of mass frame.

A photon with energy $E_\gamma > 1.3\,GeV$ is required on the recoil. Finally,
the signal yield is extracted by means of a Maximum Likelihood (ML) fit, based
on the distribution of $m_{\mathrm{ES}}$. The BR measured from a sample of about
$232 \times 10^6$ is:
\begin{equation}
 \mathcal{B}(B \to X_s \gamma) = (3.66 \pm 0.85 \pm 60)
\times 10^{-4} \, .
\end{equation}
The fit is also performed in different bins of $E_\gamma$ in order to measure
the photon energy spectrum. The results are shown in
Fig.~\ref{fig:Xsgamma_spectrum}.

\begin{figure}
  \begin{center}
  \includegraphics[width=6cm]{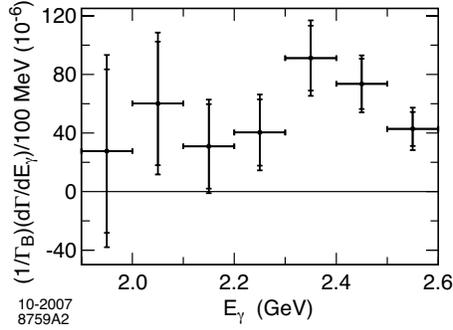}
\caption{Partial $B \to X_s \gamma$ BR in different regions of
$E_\gamma$.}
\label{fig:Xsgamma_spectrum}
\end{center}
\end{figure}

The FCNC $B \to X_s \ell^+ \ell^- (\nu \overline \nu)$ decays are also
mediated by electroweak penguin diagrams, along with electroweak box diagrams.
These processes are mainly sensitive to non-standard $Z$ couplings~\cite{BHI}.
Moreover, since the two neutrinos in the $B \to X_s \nu \overline \nu$
decays are not detected, the experimental search is also sensitive to some
exotic sources of missing energy, like light dark matter~\cite{bird} and
Unparticles~\cite{aliev}.
 
The BaBar Collaboration measured the inclusive $B \to X_s \ell^+ \ell^-$ BR by
means of a semi-inclusive approach, by looking for exclusive modes with 1 kaon 
and up to 3 pions~\cite{babar_Xsll_semi}. The result is: 
\begin{equation}
 \mathcal{B}(B \to X_s \ell^+ \ell^-) = (5.6 \pm 1.5 \pm 1.3) \times
10^{-6}\, ,
\end{equation}
and is consistent with the SM expectations. Also in this case, large systematic
uncertainties are present due to the fragmentation modeling. 

\begin{table}
\begin{tabular}{lccc}
\hline
Decay Mode & Combined & $0.1 < m^2_{ll} < 7.02\,GeV^2/c^4$ &
$m^2_{ll} > 10.24\,GeV^2/c^4$ \\
\hline
$K^+\ell^+\ell^-$ & $-0.18_{-0.18}^{+0.18}\pm0.01$ &
$-0.18_{-0.19}^{+0.19}\pm0.01$ & $-0.09_{-0.39}^{+0.36}\pm0.02$   \\
$K^{*0}\ell^+\ell^-$ & $0.02_{-0.20}^{+0.20}\pm0.02$ &
$-0.23_{-0.38}^{+0.38}\pm0.02$ & $0.17_{-0.24}^{+0.24}\pm0.02$  \\
$K^{*+}\ell^+\ell^-$ & $0.01_{-0.24}^{+0.26}\pm0.02$ &
$0.10_{-0.24}^{+0.25}\pm0.02$ & $-0.18_{-0.55}^{+0.45}\pm0.04$ \\
$K^{*}\ell^+\ell^-$ & $0.01_{-0.15}^{+0.16}\pm0.01$ &
$0.01_{-0.20}^{+0.21}\pm0.01$ & $0.09_{-0.21}^{+0.21}\pm0.02$    \\
\hline
\end{tabular}
\caption{Forward backward asymmetry for two different $m^2_{ll}$ region and in
the full range.}
\label{tab:AFB}
\end{table}

Measurements concerning the exclusive $B \to K^{(*)} \ell^+ \ell^-$ decays have
been recently updated~\cite{babar_Kll_excl}. Exclusive BR measurements need to
be compared with unclean theoretical expectations, due to the presence of large
long distance contributions~\cite{Vll}. Anyway, it is possible
to measure some interesting rate asymmetry that are sensitive to NP
effects~\cite{exclusive}, with part of the theoretical uncertainties
canceling when taking the ratios. The most important one is the forward-backward
asymmetry. A preliminary result by the \babar\ Collaboration, from a sample of
about $384 \times 10^6$ $B \overline B$ pairs, is quoted in Tab.~\ref{tab:AFB}
for two different ranges of the dilepton invariant mass $m^2_{ll}$.

Concerning the $B \to K^{(*)} \nu \overline \nu$ BRs, the Collaboration recently
presented an update of the search for the $K^+$ mode
and the first preliminary \babar\ search for the $K^{*\pm,0}$ modes. They
are performed by means of a recoil technique, using semileptonic $B \to D^{(*)}
\ell \nu$ decays for the $B_{tag}$ reconstruction. Then a $K^{(*)}$ is looked
for in the recoil, requiring that no extra track is present.

Since NP effects can strongly affect the kinematics of the $B \to K^* \nu
\overline \nu$ decay, for the first time this search has been performed by
using only the variables that are not correlated to the assumed kinematical
model.

In the $B^+ \to K^+ \nu \overline \nu$ analysis, the signal yield is extracted
with a counting technique, after a multivariate selection based on the random
forest algorithm~\cite{ranfor}. In the $B \to K^* \nu \overline \nu$ search the
yield is extracted by means of a fit to the distribution of $E_{extra}$, defined
as the sum of the energies of the neutral particles not associated either to
the $B_{tag}$ or the signal $K^{(*)}$.

The following upper limits at 90\% confidence level (CL) have
been set:
\begin{eqnarray}
 \mathcal{B}(B^+ \to K^+ \nu \overline \nu) &<& 4.2 \times 10^{-5} \, ,\\
 \mathcal{B}(B^0 \to K^{*0} \nu \overline \nu) &<& 18 \times 10^{-5} \, ,\\
 \mathcal{B}(B^+ \to K^{*+} \nu \overline \nu) &<& 9 \times 10^{-5} \, .
\end{eqnarray}
based on about $351 \times 10^6$ (for $K^+ \nu \overline \nu$) and $454 \times
10^6$ (for $K^* \nu \overline \nu$) $B \overline B$ pairs. These ULs are about a
factor 10 above the expected SM values.

\section{Leptonic $B$ decays}

Purely leptonic $B$ decays occur in the SM through annihilation diagrams, and
hence are highly suppressed. According to the SM, the only among them that
is accessible at the present $B$-Factories is the $B^+ \to \tau^+ \nu$ decay,
whose BR is expected to be at the level of $10^{-4}$. On the other hand,
enhancements are possible in some NP scenarios.

The most recent \babar\ measurements for $B^+ \to \tau^+ \nu$ have been
performed in the recoil of both hadronic and semileptonic $B_{tag}$
decays~\cite{babar_taunu}. The $\tau$ decay modes used in these analyses are
both leptonic and hadronic. The measured BR is:
\begin{eqnarray}
 \mathcal{B}(B^+ \to \tau^+ \nu) &=& (0.9 \pm 0.6 \pm 0.1)\\
 \mathcal{B}(B^+ \to \tau^+ \nu) &=& (1.8 ^{+0.9}_{-0.8} \pm 0.4)
\end{eqnarray}
for the semileptonic and hadronic analysis respectively, based on $383 \times
10^6$ $B \overline B$ pairs. They are in agreement with the expected SM values.

The \babar\ Collaboration also performed searches for the $B \to \ell^+ \ell^-$
decays, where the $e^+ e^-$, $\mu^+\mu^-$, $e^+\mu^-$, $e^+\tau^-$ and
$\mu^+\tau^-$ combinations are considered. If there is no $\tau$ in the final ,
state, a full reconstruction is performed and the signal yield is extracted with
a ML fit to the distributions of $m_{ES}$ and $\Delta E$. If a $\tau$ is
present, a hadronic recoil technique is adopted and the momentum of the other
lepton is fitted in order to extract the signal yield. The following ULs at 90\%
CL are set:
\begin{eqnarray}
 \mathcal{B}(B^0 \to e^+ e^-) < 11.3 \times 10^{-8}\\
 \mathcal{B}(B^0 \to \mu^+ \mu^-) < 5.2 \times 10^{-8}\\
 \mathcal{B}(B^0 \to e^+ \mu^-) < 9.2 \times 10^{-8}\\
 \mathcal{B}(B^0 \to e^+ \tau^-) < 2.8 \times 10^{-5}\\
 \mathcal{B}(B^0 \to \mu^+ \tau^-) < 2.2 \times 10^{-5}
\end{eqnarray}

\section{LFV $\tau$ decays}

The decays of the $\tau$ lepton can be studied at \babar\ thanks to the high
$e^+e^- \to \tau^+ \tau^-$ cross section at 10.58 $GeV$ ($\sim 0.92\,nb$). The
most recent measurements published by the Collaboration for the LFV $\tau$
decays concern $\tau^+ \to \ell^+ \omega$~\cite{babar_lomega} and $\tau^+ \to
\ell^+\ell^-\ell^+$~\cite{babar_lll}.
These decays are expected to be unaccessible in the SM
hypothesis but they can be enhanced and reach the $10^{7}$ level in some
supersymmetric scenario.

These analyses are performed by looking for a 1--3 topology: the event is
divided in two hemispheres and only one track (assumed to come from a
leptonic $\tau$ decay) is searched for in one hemisphere, while three tracks
(the signal signature) are searched for in the opposite hemisphere. Then, two
kinematical variables, the missing energy $\Delta E$ and the beam
energy constrained $\tau$ mass $m_{EC}$, are used to define a signal region. 
Finally, the observed events in the signal region are compared to the expected
yields extracted from MC simulations. The following ULs are set:
\begin{eqnarray}
 \mathcal{B}(\tau^+ \to e^+ \omega) < 10 \times 10^{-8}\\
 \mathcal{B}(\tau^+ \to \mu^+ \omega) < 11 \times 10^{-8}\\
 \mathcal{B}(\tau^+ \to e^+ e^- e^+) < 10 \times 10^{-8}\\
 \mathcal{B}(\tau^+ \to \mu^+ e^- e^+) < 11 \times 10^{-8}\\
 \mathcal{B}(\tau^+ \to \mu^- e^+ e^+) < 10 \times 10^{-8}\\
 \mathcal{B}(\tau^+ \to e^- \mu^+ \mu^+) < 11 \times 10^{-8}\\
 \mathcal{B}(\tau^+ \to e^+ \mu^- \mu^+) < 10 \times 10^{-8}\\
 \mathcal{B}(\tau^+ \to \mu^+ \mu^- \mu^+) < 11 \times 10^{-8}
\end{eqnarray}

\bibliographystyle{aipproc}

\begin{thebibliography}{9}

\bibitem{NIM}
  B.~Aubert {\it et al.}  [BABAR Collaboration],
  Nucl.\ Instrum.\ Meth.\  A {\bf 479} (2002) 1

\bibitem{geant}
  S.~Agostinelli {\it et al.}  [GEANT4 Collaboration],
  Nucl.\ Instrum.\ Meth.\  A {\bf 506} (2003) 250.
 
\bibitem{misiak} 
  M.~Misiak {\it et al.},
  Phys.\ Rev.\ Lett.\  {\bf 98} (2007) 022002
  
\bibitem{hurth}
  T.~Hurth, E.~Lunghi and W.~Porod,
  Nucl.\ Phys.\  B {\bf 704} (2005) 56
   
\bibitem{babar_Xsgamma_semi}
   B.~Aubert {\it et al.}  [BABAR Collaboration],
  Phys.\ Rev.\  D {\bf 72} (2005) 052004

\bibitem{babar_Xsgamma_reco}
  B.~Aubert {\it et al.}  [BABAR Collaboration],
  Phys.\ Rev.\  D {\bf 77} (2008) 051103

\bibitem{BHI}
  G.~Buchalla, G.~Hiller and G.~Isidori,
  Phys.\ Rev.\  D {\bf 63} (2001) 014015.

\bibitem{bird}
  C.~Bird, P.~Jackson, R.~Kowalewski and M.~Pospelov,
  Phys.\ Rev.\ Lett.\  {\bf 93} (2004) 201803.

\bibitem{aliev}
  T.~M.~Aliev, A.~S.~Cornell and N.~Gaur,
  JHEP {\bf 0707} (2007) 072
  
\bibitem{babar_Xsll_semi}
  B.~Aubert {\it et al.}  [BABAR Collaboration],
  Phys.\ Rev.\ Lett.\  {\bf 93} (2004) 081802
  
\bibitem{babar_Xsll_ACP}
  B.~Aubert {\it et al.}  [BABAR Collaboration],
  arXiv:0805.4796 [hep-ex].
\bibitem{babar_Kll_excl}
  B.~Aubert {\it et al.}  [BABAR Collaboration],
  arXiv:0807.4119 [hep-ex].
  
\bibitem{Vll}
  M.~Beneke, T.~Feldmann and D.~Seidel,
  Nucl.\ Phys.\  B {\bf 612} (2001) 25
    
\bibitem{exclusive}    
  T.~Feldmann and J.~Matias,
  JHEP {\bf 0301} (2003) 074
  C.~Bobeth, G.~Hiller and G.~Piranishvili,
  JHEP {\bf 0807} (2008) 106

\bibitem{ranfor}
  L.~Breiman, Machine Learning {\bf 45}, 5-32 (2001).

\bibitem{babar_taunu}
  B.~Aubert {\it et al.}  [BABAR Collaboration],
  Phys.\ Rev.\  D {\bf 76} (2007) 052002
  B.~Aubert {\it et al.}  [BABAR Collaboration],
  Phys.\ Rev.\  D {\bf 77} (2008) 011107
  
\bibitem{babar_lomega}
  B.~Aubert {\it et al.}  [BABAR Collaboration],
  Phys.\ Rev.\ Lett.\  {\bf 100} (2008) 071802
  
\bibitem{babar_lll}
  B.~Aubert {\it et al.}  [BABAR Collaboration],
  Phys.\ Rev.\ Lett.\  {\bf 99} (2007) 251803
  
\end{thebibliography}

\end{document}